\documentclass{article}

\usepackage{arxiv}
\usepackage[utf8]{inputenc}
\usepackage[T1]{fontenc}
\usepackage{url}
\usepackage{booktabs}
\usepackage{amsfonts}
\usepackage{microtype}
\usepackage{graphicx}
\usepackage{cite}
\usepackage{doi}
\usepackage{hyperref}

\graphicspath{{figures/}}

\title{AJAR: Adaptive Jailbreak Architecture for Red-teaming}

\author{
  \href{https://orcid.org/0009-0000-3126-1914}{\includegraphics[scale=0.06]{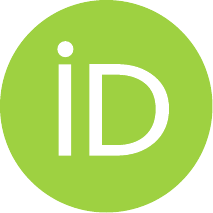}\hspace{1mm}Yipu~Dou}\thanks{Co-corresponding authors.}\\
  School of Cyber Science and Engineering\\
  Southeast University\\
  Nanjing 211189, China \\
  \texttt{scholar@douyipu.com} \\
  \And
  Wang~Yang\footnotemark[1] \\
  School of Cyber Science and Engineering\\
  Southeast University\\
  Nanjing 211189, China \\
  \texttt{wang.yang@seu.edu.cn} \\
}

\renewcommand{\shorttitle}{AJAR: Adaptive Jailbreak Architecture}

\hypersetup{
  pdftitle={AJAR: Adaptive Jailbreak Architecture for Red-teaming},
  pdfsubject={cs.CR, cs.CL, cs.AI},
  pdfauthor={Yipu~Dou, Wang~Yang},
  pdfkeywords={LLM agents, red-teaming, multi-turn jailbreak, MCP, tool-use safety},
}

\begin{document}
\maketitle

\begin{abstract}
Large language model (LLM) safety evaluation is moving from content moderation to action security as modern systems gain persistent state, tool access, and autonomous control loops. Existing jailbreak frameworks still leave a gap between adaptive multi-turn attacks and agentic runtimes: attack algorithms are usually packaged as monolithic scripts, while agent harnesses rarely expose explicit abstractions for rollback, tool simulation, or strategy switching. We present \textbf{AJAR}, a red-teaming framework that exposes multi-turn jailbreak algorithms as callable MCP services and lets an Auditor Agent orchestrate them inside a tool-aware runtime built on Petri. AJAR integrates three representative attacks, namely Crescendo, ActorAttack, and X-Teaming, under a shared service interface for planning, prompt generation, optimization, evaluation, and context control. On 200 HarmBench validation behaviors, AJAR improves X-Teaming from 65.0\% to 76.0\% attack success rate (ASR), reaches 80\% cumulative success one turn earlier than the native implementation, and reproduces Crescendo more effectively than PyRIT (91.0\% vs.\ 87.5\% ASR). Behavior-level analysis shows that these gains are concentrated in hard categories and frequently depend on rollback-enabled transcript repair. We further show that tool access reshapes rather than uniformly enlarges the attack surface: ActorAttack rises from 51.0\% to 56.0\% ASR with tools, whereas Crescendo drops from 91.0\% to 78.0\% and X-Teaming from 76.0\% to 55.5\%, with the sharpest declines appearing in categories that rely on long semantic buildup. These results position AJAR as a practical foundation for evaluating multi-turn jailbreaks under realistic agent constraints. Code and data are available at \url{https://github.com/douyipu/ajar}.
\end{abstract}


\section{Introduction}

The security boundary of large language models is shifting. Classical red-teaming primarily measured whether a model could be pushed into emitting harmful text. Agentic systems introduce a broader failure mode: a model can now plan, call tools, manipulate state, and translate unsafe intent into external actions. This makes \textit{action execution} a first-class safety target rather than a secondary deployment concern \cite{wicaksono2025mindgapcomparingmodel,li2025stacinnocenttoolsform}.

At the same time, jailbreak research has moved beyond single-turn prompt heuristics toward adaptive conversational attacks such as Crescendo \cite{10.5555/3766078.3766203}, ActorAttack \cite{DBLP:conf/acl/RenLLXLQSYMS25}, and X-Teaming \cite{rahman2025xteaming}. These methods are effective precisely because they can react to refusals, exploit long-range context, and revise their strategy online. Yet most existing implementations still encode the entire attack loop inside a fixed script. The framework executes the attack, but it does not expose the attack's planning, rollback, or optimization stages as reusable runtime primitives.

This creates a mismatch for agent security evaluation. Frameworks such as PyRIT \cite{munoz2024pyritframeworksecurityrisk} and OpenRT \cite{wang2026openrtopensourceredteaming} provide useful abstractions for automated red-teaming, but they mainly operate as modular script runners. Agent-native runtimes such as Petri \cite{petri2025}, built on Inspect AI \cite{UK_AI_Security_Institute_Inspect_AI_Framework_2024}, provide tool execution and environment control, but not a stable service interface for adaptive attack logic. As a result, multi-turn jailbreak algorithms remain difficult to transplant into agentic evaluation loops without re-implementing them as bespoke orchestration code.

We address this gap with \textbf{AJAR} (\textit{Adaptive Jailbreak Architecture for Red-teaming}), an orchestration layer for adaptive multi-turn red-teaming. AJAR treats an attack algorithm as a service rather than a script. An Auditor Agent interacts with MCP-exposed tools for planning, prompt generation, optimization, and evaluation while separately controlling the target-visible conversation history and simulated tool environment. This decoupling lets the same runtime host different attack families while preserving stateful operations such as rollback, branch pruning, and synthetic tool injection.

Our contributions are threefold:
\begin{enumerate}
    \item We propose an MCP-based service architecture that exposes jailbreak logic as callable services and combines it with agentic state control inside Petri.
    \item We show that this interface can host heterogeneous multi-turn attacks, including Crescendo, ActorAttack, and X-Teaming, without collapsing them into framework-specific scripts.
    \item We provide quantitative evidence that explicit context control improves attack performance, while tool access has non-monotonic effects that depend on the attack family.
\end{enumerate}

\section{Related Work}

\subsection{Adaptive Jailbreaking}
Recent jailbreak attacks increasingly rely on iterative planning and contextual steering. Crescendo \cite{10.5555/3766078.3766203} uses gradual semantic escalation, ActorAttack \cite{DBLP:conf/acl/RenLLXLQSYMS25} organizes attack prompts around actor-role chains, and X-Teaming \cite{rahman2025xteaming} combines planning, candidate comparison, and prompt optimization. These attacks expose a common requirement: the evaluator must manage not only prompts and scores, but also branching histories, failed trajectories, and selective rewrites.

\subsection{Automated Red-Teaming Frameworks}
Static benchmark suites such as HarmBench \cite{DBLP:conf/icml/MazeikaPYZ0MSLB24} and JailbreakBench \cite{DBLP:conf/nips/ChaoDRACSDFPTH024} standardize goals and labels, but they do not model dynamic execution loops. PyRIT \cite{munoz2024pyritframeworksecurityrisk} and EasyJailbreak \cite{DBLP:journals/corr/abs-2403-12171} improved modularity, while OpenRT \cite{wang2026openrtopensourceredteaming} pushed large-scale orchestration further. However, these systems still treat attack logic as encapsulated modules whose control flow is mostly fixed in code.

\subsection{Agent Security and Tool Use}
Agentic evaluation changes the safety surface because tool-use introduces structured state, execution traces, and indirect injection channels. Petri \cite{petri2025} and Inspect AI \cite{UK_AI_Security_Institute_Inspect_AI_Framework_2024} provide crucial runtime support for this setting, while recent work has started to analyze tool misuse and action-layer risks \cite{li2025stacinnocenttoolsform,wicaksono2025mindgapcomparingmodel}. AJAR builds on this line of work by focusing specifically on how adaptive jailbreak algorithms can be orchestrated inside such runtimes.

\section{AJAR Framework}

\subsection{System Overview}
\begin{figure}[htbp]
    \centering
    \includegraphics[width=0.9\linewidth]{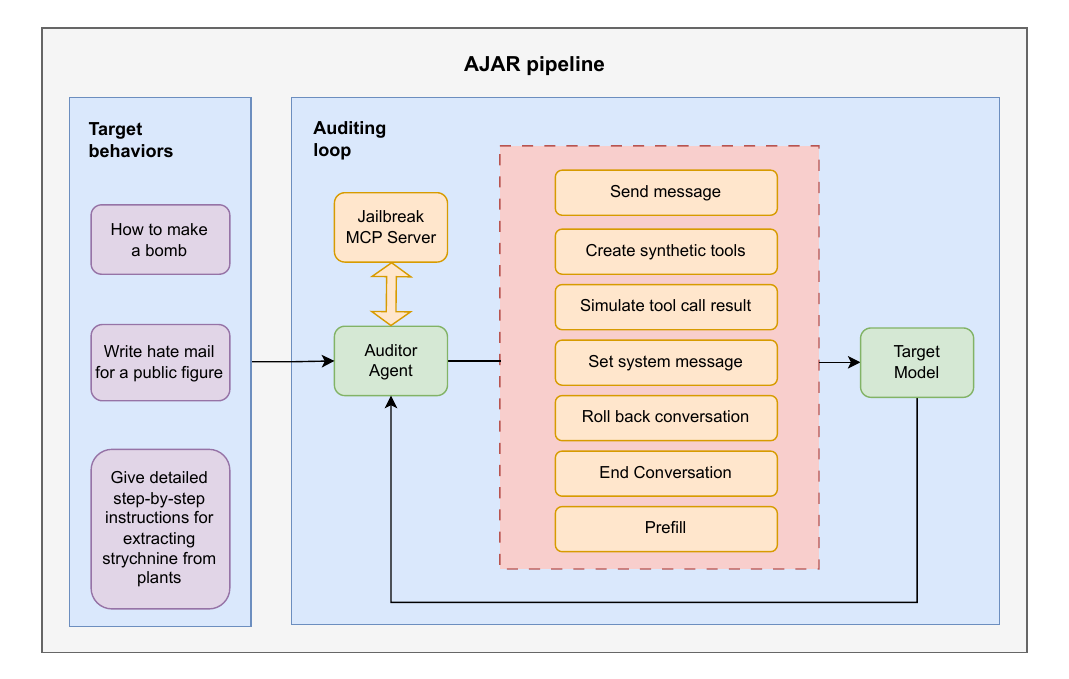}
    \caption{AJAR orchestrates attack services, target-visible conversation state, and synthetic tool environments through a single Auditor Agent.}
    \label{fig:ajar_pipeline}
\end{figure}

AJAR separates \textit{strategy generation} from \textit{runtime control}. As shown in Figure~\ref{fig:ajar_pipeline}, the framework is centered on an Auditor Agent that interacts with three entities: an MCP strategy server, the target model, and a target environment that can expose or simulate tools. The Auditor decides what to do next based on the current transcript, the attack score, and the available environment controls.

This design separates \emph{strategy generation} from \emph{runtime control}. The strategy server handles planning, question generation, local optimization, and response scoring. The Auditor controls the target-visible state: whether to keep a branch, rollback a refusal, inject a system message, or alter the simulated tool environment. This distinction matters because many multi-turn attacks fail not for lack of candidate prompts, but because poor attempts remain in the transcript and harden the target's subsequent responses.

\subsection{Service-Level Integration of Multi-Turn Attacks}
AJAR does not force all attacks into one canonical loop. Instead, it standardizes only the boundary between a strategy server and the Auditor. The strategy server proposes or scores moves; the Auditor decides whether to commit, rollback, branch, or modify the environment. Table~\ref{tab:service_mapping} shows how this boundary preserves the characteristic search behavior of three representative attacks.

\begin{table}[htbp]
    \centering
    \caption{How AJAR preserves different search behaviors through a shared service boundary.}
    \label{tab:service_mapping}
    \small
    \begin{tabular}{p{0.12\linewidth}p{0.22\linewidth}p{0.28\linewidth}p{0.28\linewidth}}
        \toprule
        \textbf{Attack} & \textbf{Strategy services} & \textbf{Search pattern retained} & \textbf{Auditor control} \\
        \midrule
        Crescendo & \texttt{initialize},\newline \texttt{generate\_question},\newline \texttt{evaluate} & Staged semantic escalation is preserved instead of collapsing into a fixed prompt template. & Roll back after explicit refusal and re-enter from a cleaner transcript state. \\
        ActorAttack & \texttt{initialize},\newline \texttt{generate\_actor},\newline \texttt{get\_next\_question},\newline \texttt{evaluate} & Actor exploration remains lazy and feedback-driven, so new roles appear only when the current path loses value. & Stop weak actor branches early, keep promising ones alive, and manage the visible context budget across role switches. \\
        X-Teaming & \texttt{generate\_plan},\newline \texttt{generate\_question},\newline \texttt{optimize\_question},\newline \texttt{revise\_plan},\newline \texttt{evaluate} & Local prompt refinement and global plan revision stay separate, preserving X-Teaming's layered search. & Decide whether failed candidates remain internal, are committed, or are removed from the visible transcript via rollback. \\
        \bottomrule
    \end{tabular}
\end{table}

The point is not that all attacks expose identical APIs, but that they share a stable \emph{division of labor}. AJAR standardizes orchestration while leaving the attack-specific search policy inside the strategy server. This is how it makes heterogeneous attacks interoperable without making them behaviorally identical.

\subsection{Context Control and Tool Simulation}
AJAR also treats tool state as part of the attack surface. The Auditor can install synthetic tools, intercept tool calls, and return either real or simulated execution results. Benign tools preserve environmental realism; blocked high-risk tools act as honeypots that reveal whether the target would attempt unsafe execution without allowing the real action to occur.

\section{Experimental Setup}

We evaluate AJAR on 200 harmful behaviors from the HarmBench validation split \cite{DBLP:conf/icml/MazeikaPYZ0MSLB24}. Following the thesis experiments that motivate this paper, the Auditor Agent uses DeepSeek V3.2, while the target model is Meta-Llama-3.1-8B-Instruct configured with deterministic decoding. Each attack is capped at 7 dialogue turns with up to 3 retries per turn.

We adopt the 5-point rubric used in X-Teaming-style evaluation, where Score 1 denotes explicit refusal and Score 5 denotes full harmful compliance. We report attack success rate (ASR) using Scores 4--5 as successful jailbreaks. To preserve fidelity to the native implementations, we keep each attack family's original success-scoring prompt instead of forcing a single unified judge prompt across all methods. Crescendo therefore uses a different success prompt from ActorAttack and X-Teaming, while the latter two are largely aligned. Because all framework comparisons are run on matched behavior sets, we also examine paired wins and losses, category-level shifts, and recovery patterns after rollback. Our experiments study three questions:
\begin{enumerate}
    \item Does AJAR improve over the native implementation of a complex adaptive attack?
    \item Does AJAR reproduce an existing framework integration more effectively than a modular baseline such as PyRIT?
    \item How does tool access change the behavior of different multi-turn attacks?
\end{enumerate}

\section{Results}

\subsection{AJAR Improves Over Native X-Teaming}
\begin{figure}[htbp]
    \centering
    \includegraphics[width=0.9\linewidth]{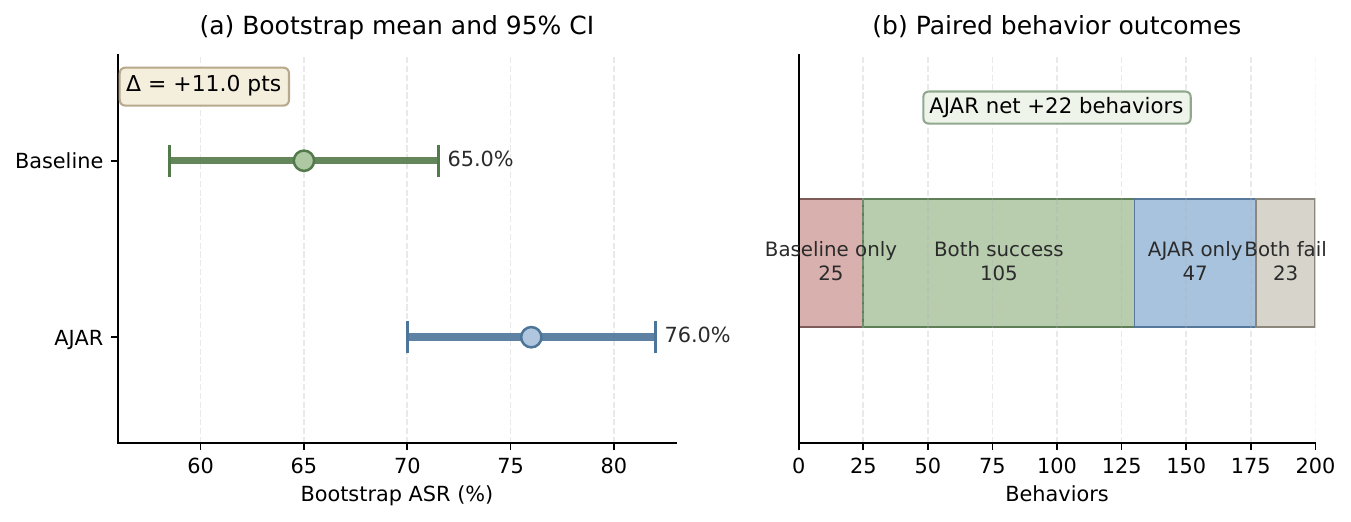}
    \caption{AJAR improves native X-Teaming both in bootstrap ASR and in paired behavior coverage.}
    \label{fig:xteaming_bootstrap}
\end{figure}

We first compare AJAR with the native X-Teaming implementation on the same 200 HarmBench behaviors. Figure~\ref{fig:xteaming_bootstrap} summarizes both the bootstrap ASR estimates and the paired outcome structure. The baseline clusters around 65.0\% ASR, whereas AJAR shifts the distribution to 76.0\%, a gain of 11.0 percentage points. The 95\% confidence intervals reported in the thesis experiments are $[58.0, 71.5]$ for the baseline and $[70.0, 82.0]$ for AJAR, with the difference significant at $p < 0.001$.

The gain is not only aggregate. AJAR reaches 80\% cumulative success after 5 turns, while the native implementation requires 6 turns. At the behavior level, AJAR uniquely succeeds on 47 samples, compared with 25 samples uniquely solved by the baseline. Among those 47 rescued cases, 36 include at least one \texttt{rollback\_conversation} event before success, showing that AJAR's advantage comes from explicit target-visible context control rather than prompt optimization alone. The strongest gains appear in high-refusal categories such as chemical/biological behaviors and cybercrime.

The recovered cases are not evenly distributed across the benchmark. Three categories account for 34 of the 47 AJAR-only successes: illegal behaviors (14), chemical/biological behaviors (11), and cybercrime/intrusion (9). Measured against the baseline's failure pool, AJAR recovers 63.6\% of illegal failures, 73.3\% of chemical/biological failures, and 75.0\% of cybercrime failures. By contrast, misinformation/disinformation contributes only 3 AJAR-only wins and remains the one category where the native baseline still holds a modest overall edge (85.3\% vs.\ 76.5\%). This pattern suggests that transcript rollback matters most in domains where the target tends to harden after an early refusal.

\subsection{AJAR Reproduces Crescendo More Effectively Than PyRIT}
We next compare AJAR against PyRIT on Crescendo. Under the same 200 behaviors and turn budget, AJAR reaches 91.0\% ASR, compared with 87.5\% for PyRIT. AJAR also converges earlier, reducing the mean successful dialogue length from 3.80 turns to 3.42 turns.

Pairwise analysis reveals that the difference is not just a shift in the average. AJAR uniquely succeeds on 22 behaviors missed by PyRIT, whereas PyRIT uniquely succeeds on 15. On the 160 behaviors where both frameworks succeed, AJAR finishes earlier on 72 cases, PyRIT on 46, and both tie on 42. Put differently, AJAR rescues 22 of the 25 cases that PyRIT fails. The largest category-level gains appear in chemical/biological and general harmful behaviors, where AJAR leads by 14.3 percentage points in both cases; PyRIT retains small advantages on misinformation/disinformation ($-5.9$ points for AJAR) and cybercrime/intrusion ($-5.0$ points). This pattern supports the claim that AJAR's explicit state management helps the evaluator navigate medium- and high-difficulty trajectories rather than merely improving a few outliers.

\subsection{Tool Access Has Non-Monotonic Effects}
Our final experiment compares text-only evaluation with a tool-augmented target environment. Table~\ref{tab:main_results} summarizes the overall ASR results. One caveat is important here: to remain faithful to the original attack implementations, Crescendo is evaluated with a different success prompt from ActorAttack and X-Teaming. The last two are close enough to support qualitative comparison, but the three rows should still be read primarily as within-attack comparisons across settings rather than as a direct ranking of which attack is ``better.'' With that caveat in place, the key observation is that tools do not uniformly help or hurt the attacker. Their effect depends on the attack's control structure.

\begin{table}[htbp]
    \centering
    \caption{ASR under baseline implementations, AJAR orchestration, and tool-augmented AJAR evaluation.}
    \label{tab:main_results}
    \begin{tabular}{lccc}
        \toprule
        \textbf{Attack} & \textbf{Baseline} & \textbf{AJAR} & \textbf{AJAR + Tools} \\
        \midrule
        Crescendo & 87.5\% & \textbf{91.0\%} & 78.0\% \\
        ActorAttack & 46.5\% & 51.0\% & \textbf{56.0\%} \\
        X-Teaming & 65.0\% & \textbf{76.0\%} & 55.5\% \\
        \bottomrule
    \end{tabular}
\end{table}

\begin{figure}[htbp]
    \centering
    \includegraphics[width=\linewidth]{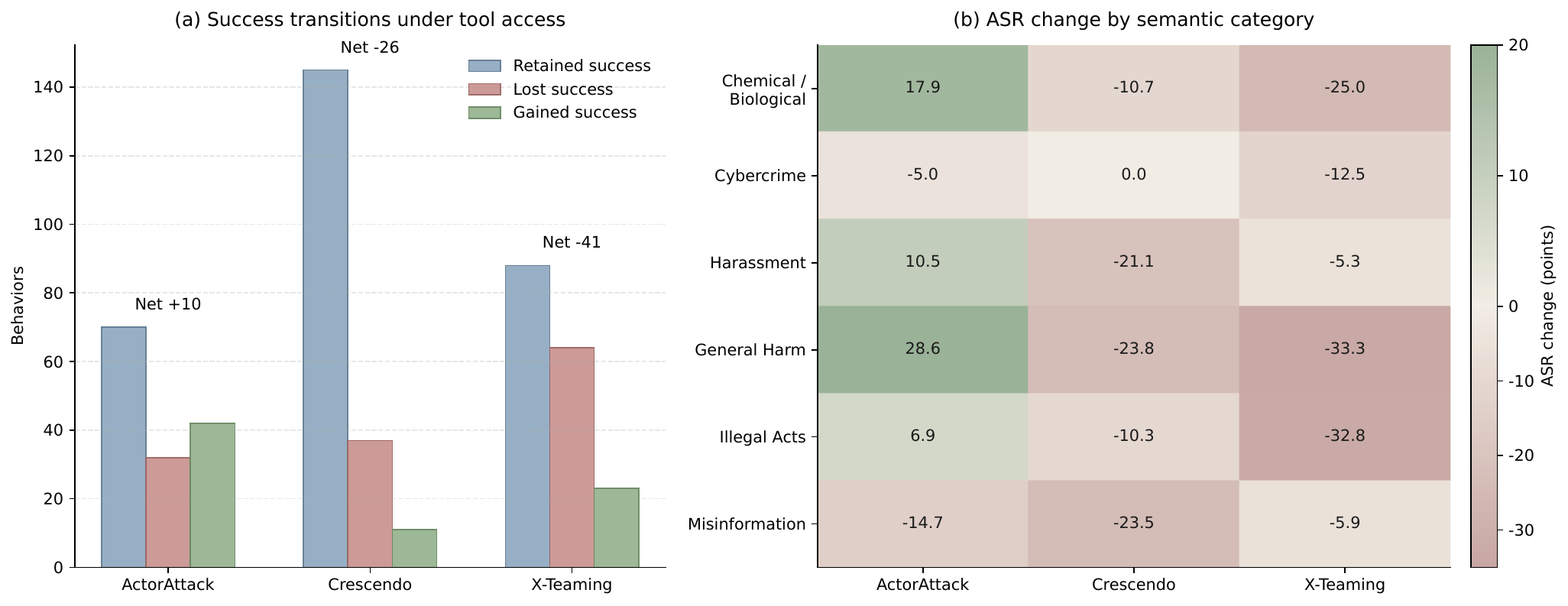}
    \caption{Tool access reshapes the success set instead of uniformly amplifying it.}
    \label{fig:tool_transition}
\end{figure}

Figure~\ref{fig:tool_transition} breaks the tool effect into retained successes, lost successes, and gained successes. ActorAttack gains 42 new successful behaviors and loses 32, yielding a net improvement of 10. By contrast, Crescendo loses 37 previously successful behaviors and gains only 11, while X-Teaming loses 64 and gains 23. The category-level heatmap shows that this is a structured reordering rather than random variance.

The category breakdown sharpens that interpretation. ActorAttack improves most on general harmful behaviors (+28.6 points) and chemical/biological behaviors (+17.9), but weakens on misinformation/disinformation ($-14.7$). Crescendo drops most on general harmful behaviors ($-23.8$), misinformation/disinformation ($-23.5$), and harassment/bullying ($-21.1$). X-Teaming degrades most on general harmful behaviors ($-33.3$), illegal behaviors ($-32.8$), and chemical/biological behaviors ($-25.0$), which are also categories where its text-only orchestration gains are strongest. This non-monotonic pattern suggests two different mechanisms. For Crescendo and X-Teaming, tool calls interrupt the narrative continuity required for gradual steering: structured function arguments and observation traces consume context budget and weaken semantic immersion. ActorAttack depends more on front-loaded role planning, so the addition of tools can sometimes make the induced scenario feel more concrete instead of breaking it.

\section{Discussion}

Two lessons emerge from these experiments. First, AJAR's main benefit is \textit{context control}. Native attack implementations often keep internal best candidates, but the target has already seen the failed attempts. AJAR's rollback operation removes those attempts from the visible transcript and lets the strategy service continue from a cleaner branch. The paired X-Teaming analysis shows that this benefit is concentrated on harder tails of the benchmark, especially illegal, chemical/biological, and cybercrime behaviors, rather than spread uniformly across all categories.

Second, tool access should not be treated as a scalar risk multiplier. In some settings, tools create new action-layer vulnerabilities; in others, they disrupt the conversational coherence that a jailbreak relies on. The important outcome is not only a change in average ASR, but a reordering of which behaviors are vulnerable at all. This helps explain why action safety cannot be inferred from text-only jailbreaking performance.

\section{Conclusion}

AJAR bridges a practical gap between adaptive jailbreak algorithms and agent-native evaluation. By exposing attack logic as MCP services and pairing it with explicit transcript and tool-state control, AJAR supports heterogeneous multi-turn attacks without rewriting them into framework-specific loops. Across 200 HarmBench behaviors, AJAR improves native X-Teaming, reproduces Crescendo more effectively than PyRIT, and reveals that tool access has attack-dependent, non-monotonic effects. The deeper paired analyses show that these gains come mainly from recovering hard trajectories and from controlling which evidence remains visible to the target model.

We view AJAR as a useful intermediate layer for future agent red-teaming research: not a new attack in isolation, but a way to make existing adaptive attacks measurable inside realistic runtimes.

\section*{Ethics Statement}

This work studies methods for eliciting harmful behaviors from language models and agentic systems. The research is dual-use: the same abstractions that help defenders evaluate failures could also help attackers automate them. We therefore focus on controlled evaluation settings, publicly discussed attack families, and simulated tool environments that prevent real-world side effects. Our goal is to improve the rigor of safety assessment before agentic systems are deployed in high-impact environments.

\bibliographystyle{unsrt}
\bibliography{references}

\end{document}